\documentclass{jpsj3}
\usepackage{txfonts}
\usepackage{color}

\title{Pronounced $-$ Log $T$ Divergence in Specific Heat of Nonmetallic CeOBiS$_{2}$: A Mother Phase of BiS$_{2}$-Based Superconductor}

\author{Ryuji Higashinaka$^{1,}$\thanks{E-mail address: higashin@tmu.ac.jp}, Takuya Asano$^{1}$, Takuya Nakashima$^{1}$, Kengo Fushiya$^{1}$, Yoshikazu Mizuguchi$^{2}$, Osuke Miura$^{2}$, Tatsuma D. Matsuda$^{1}$, and Yuji Aoki$^{1}$}
\inst{$^1$Department of Physics, Tokyo Metropolitan University, Hachioji, Tokyo 192-0397, Japan\\
$^2$Department of Electrical and Electronic Engineering, Tokyo Metropolitan University, Hachioji, Tokyo 192-0397, Japan
}
\abst{
The low-temperature properties of CeOBiS$_{2}$ single crystals are studied by electrical resistivity, magnetization, and specific heat measurements.
Ce 4$f$-electrons are found to be in a well-localized state split by crystalline-electric-field (CEF) effects.
The CEF ground state is a pure $J_{z} = \pm 1/2$ doublet, and excited doublets are located far above.
At low temperatures in zero field, we observe pronounced $- \log T$ divergence in the specific heat, revealing the presence of quantum critical fluctuations of 4$f$ magnetic moments near a quantum critical point (QCP).
Considering that CeOBiS$_{2}$ is a nonmetal, this phenomenon cannot be attributed to the competition between Kondo and the Ruderman-Kittel-Kasuya-Yosida (RKKY) interactions as in numerous $f$-electron-based strongly correlated metals, indicating an unconventional mechanism.
We suggest that CeOBiS$_{2}$ is the first material found to be located at a QCP among geometrically frustrated nonmetallic magnets.
}
\date{\today}
\begin{document}
\maketitle
\begin{figure}
\begin{center}
\includegraphics[width=0.7\linewidth]{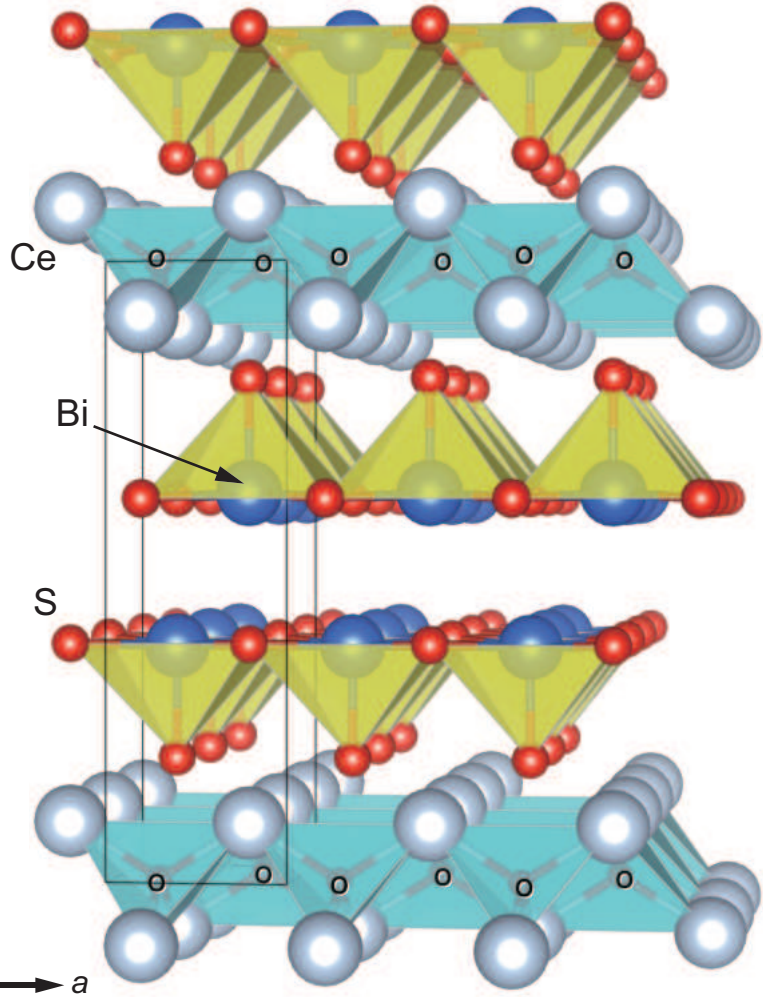}
\end{center}
\caption{(Color online) Tetragonal crystal structure ($P4/nmm$) of CeOBiS$_{2}$, consisting of an alternating stacking of BiS$_{2}$ and CeO layers (depicted using VESTA~\cite{VESTA}).
A unit cell is indicated by solid lines.
}
\label{Crystal_Structure}
\end{figure}
\begin{figure}
\begin{center}
\includegraphics[width=\linewidth]{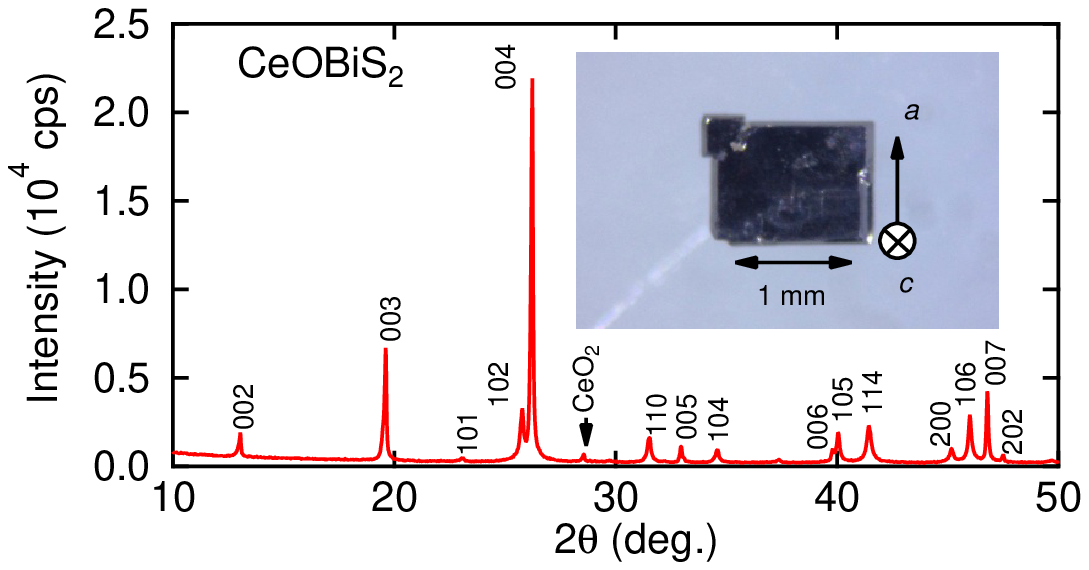}
\end{center}
\caption{(Color online) X-ray powder diffraction pattern for crushed single crystals of CeOBiS$_{2}$ using a Cu-K$\alpha 1$ radiation. 
The Miller indices of the tetragonal symmetry $P4/nmm$ for main peaks are shown.
There are a few peaks attributed to the remaining CeO$_{2}$.
(inset) Picture of CeOBiS$_{2}$ single crystal.
}
\label{XRD}
\end{figure}

%
The $- \log T$ divergence of specific heat has been observed in numerous $f$-electron-based strongly correlated electron systems~\cite{Lohneysen_2007, Stewart_2001}.
The $- \log T$ behavior is considered to be a hallmark of non-Fermi-liquid (NFL) behavior realized around a quantum critical point (QCP) or a $T=0$ phase transition, where quantum fluctuations of magnetic moments dominate.
Since the realization of such a QCP requires the balance between Kondo and the Ruderman-Kittel-Kasuya-Yosida (RKKY) interactions, fine tuning of the relevant control parameter by chemical doping is needed in many cases.
Since chemical doping inevitably introduces randomness into a crystal, a QCP realized in a pure compound is beneficial for research to clarify the mechanism.
However, the numbers of such examples (including CeNi$_{2}$Ge$_{2}$~\cite{Steglich_1996} and $\beta$-YbAlB$_{4}$~\cite{Matsumoto_2011}) are limited.
In this letter, we report the finding of $- \log T$ divergence of the specific heat in nondoped CeOBiS$_2$.
Since this material is a nonmetal, the Kondo effect (a Fermi surface effect) cannot be realized in principle.
Therefore, an unconventional mechanism is necessary to account for this behavior.
\\
\quad 
CeOBiS$_2$ is a mother phase of a recently found BiS$_{2}$-based superconductor~\cite{Mizuguchi_PRB_12, Mizuguchi_JPSJ_12, Singh_JACS_12, Demura_condmat_12}.
As shown in Fig. \ref{Crystal_Structure}, the crystal structure consists of an alternating stacking of BiS$_{2}$ and CeO layers along the $c$-direction.
In CeO$_{1-x}$F$_{x}$BiS$_{2}$, the partial replacement of O by F provides electron doping into the BiS$_{2}$ layers and induces superconductivity with the transition temperature $T_{\rm c}$ of 2.3 K (under applied pressure, it increases to as high as 6.7 K \cite{Wolowiec_PRB_13}).
In view of the two-dimensional (2D) nature of the crystal structure, it is intriguing to compare the physical properties of CeO$_{1-x}$F$_{x}$BiS$_{2}$ with those of cuprates \cite{Cuprate_Review} and Fe-based superconductors \cite{Kamihara_JACS_08, Pnictide_Review}.
Single-crystal studies have shown that CeO$_{1-x}$F$_{x}$BiS$_{2}$ has strongly anisotropic superconducting properties \cite{Nagao_SSC_14}.
In addition, the possible coexistence of superconductivity and Ce $4f$-electron ferromagnetism in this system has been pointed out recently using polycrystalline samples~\cite{Xing_PRB_12, Demura_condmat_13}.
These findings have motivated us to investigate in detail the $4f$-electron magnetism using single crystals.
In this study, we investigate the physical properties of undoped CeOBiS$_{2}$ single crystals.
This is the first report on the magnetism of $4f$ electrons in the series of $Ln$O$_{1-x}$F$_{x}$BiS$_{2}$ ($Ln$: rare earth) superconductors using single crystals.
\\
%
%
\quad Single crystals of CeOBiS$_{2}$ were prepared by the CsCl flux method using Ce chips (99.9\%) and powders of CeO$_{2}$ (99.99\%), Bi$_{2}$S$_{3}$ (99.999\%), S (99.9\%), and CsCl (99.9\%). 
The excess CsCl flux was removed using H$_{2}$O. 
The single crystals are plate-shaped parallel to the $c$-plane with typical dimensions of 1 $\times$ 1 $\times$ 0.2 mm$^{3}$ (see inset of Fig.~\ref{XRD}).
The powder X-ray diffraction spectra shown in Fig.~\ref{XRD} confirm the LaOBiS$_{2}$-type structure with $P4/nmm$ symmetry. 
For the determination of detailed structural parameters, single-crystal X-ray analysis was performed using a Rigaku Mercury diffractometer with graphite monochromated Mo-K$\alpha$ radiation. 
A selected small single crystal with dimensions of about 0.10 $\times$ 0.10 $\times$ 0.05 mm$^{3}$ was mounted on a glass fiber with epoxy. 
Structural parameters refined (with the occupation ratio for each site fixed to 1) using the program SHELX-97~\cite{SHLEX-97} are summarized in Table I. 
The lattice parameters agree well with the reported values~\cite{Xing_PRB_12}.
DC magnetization ($M$) measurements were carried out in a Magnetic Property Measurement System (MPMS; Quantum Design (QD)) down to 2 K and up to 7 T. 
Specific heat ($C$) measurements were performed using a quasi-adiabatic heat-pulse method with a Physical Property Measurement System (PPMS; QD) and a dilution refrigerator down to 0.2 K up to 8 T. 
Considering that the measured single crystal includes nonmagnetic CeO$_{2}$ and CsCl phases (see Fig.~\ref{XRD}), a sample mass correction is made by multiplying the measured values of $M$ and $C$ by a correction factor of $p$ = 1.12. 
$p$ is determined by crystalline-electric-field (CEF) model fitting to the $M(T)$ data (see below).
Electrical resistivity was measured by a standard AC four-probe technique. 
\\
\begin{table}[tb]
\caption{Atomic coordinates and thermal parameters of CeOBiS$_2$ at room temperature determined by single-crystal X-ray measurements. $R$ and $wR$ are reliability factors and $B_{\rm eq}$ is the equivalent isotropic atomic displacement parameter. Standard deviations in the positions of the least significant digits are given in parentheses.}
\label{t1}
\begin{center}
\begin{tabular}{lrlcccl}
\hline
\multicolumn{2}{c}{$P4/nmm$ ($\sharp$129)} & \multicolumn{5}{c}{$a$ $=$ 3.984(6) $\AA$, $c$ $=$ 13.49(2) $\AA$, $V$ $= $ 214.2(6) $\AA^3$}\\
\multicolumn{2}{c}{(origin choice 2)} & \multicolumn{1}{c}{ } & \multicolumn{3}{c}{Position}\\
 \cline{4-6}
Atom & site & & $x$ & $y$ & $z$ & $B_{\rm eq}\ (\AA^2)$\\
\hline
Ce & $2c$ & ($4mm$) & 1/4 & 1/4 & 0.0919(2) & 0.35(7) \\
Bi & $2c$ & ($4mm$) & 1/4 & 1/4 & 0.6272(2) & 0.76(7) \\
S(1) & $2c$ & ($4mm$) & 1/4 & 1/4 & 0.3817(11) & 1.2(3) \\
S(2) & $2c$ & ($4mm$) & 1/4 & 1/4 & 0.8126(9) & 0.9(3) \\
O & $2b$ & (${\bar 4}m2$) & 3/4 & 1/4 & 0 & 0.4(6) \\
\hline
\multicolumn{3}{c}{$R$ $=$ 8.39 $\%$, $wR$ $=$ 13.75 $\%$}\\
\end{tabular}
\end{center}
\end{table}
%
%
\begin{figure}
\begin{center}
\includegraphics[width=\linewidth]{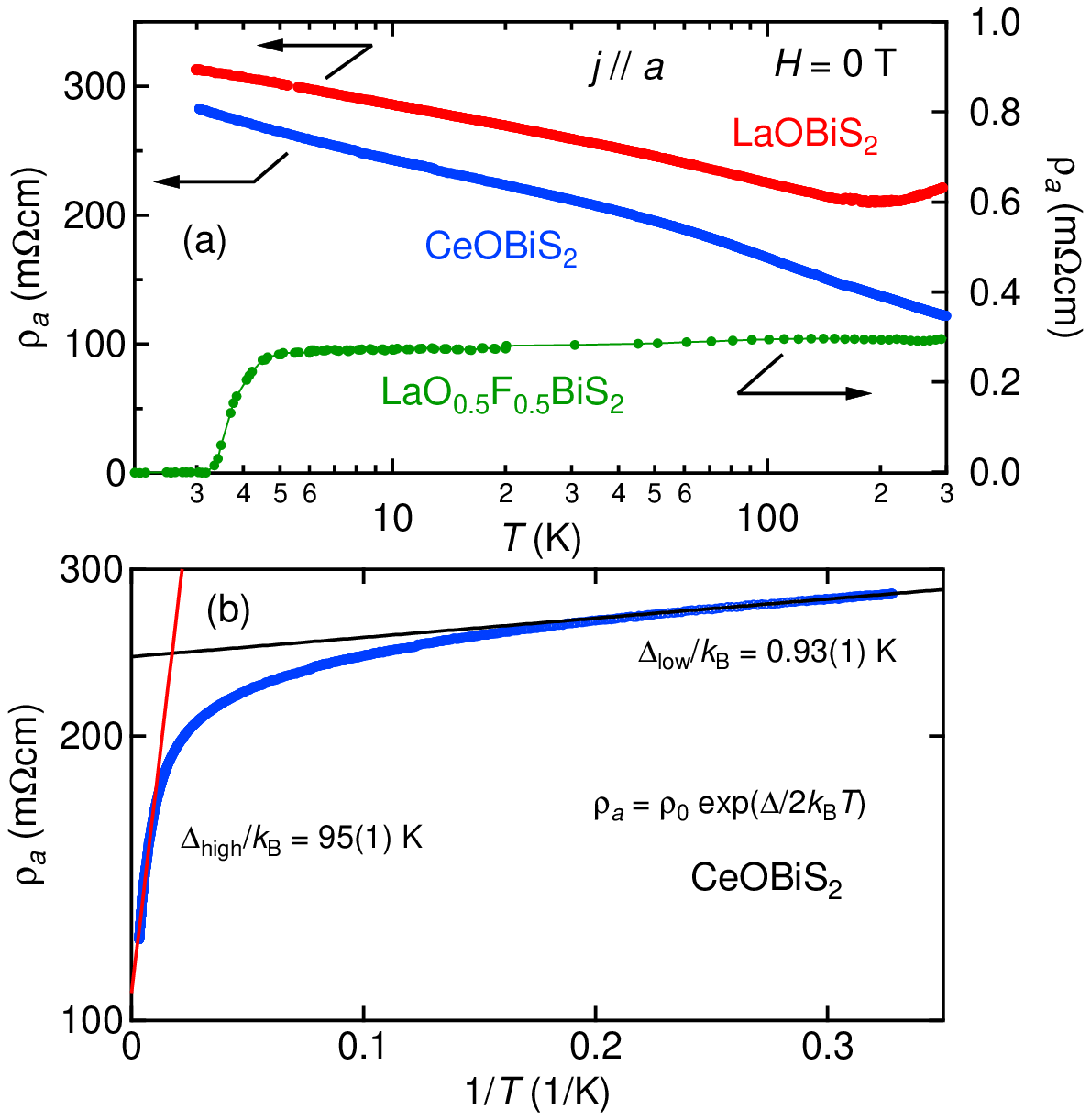}
\end{center}
\caption{(Color online) (a) Temperature dependence of the in-plane resistivity $\rho_{a}$ for single crystals of $Ln$OBiS$_{2}$ ($Ln$ = La, Ce) and  LaO$_{0.5}$F$_{0.5}$BiS$_{2}$ at 0 T.
(b) $1/T$ dependence of $\log \rho_{a}$ for CeOBiS$_{2}$.
The solid lines are simple activation-type fitting curves in the low- and high-temperature regions.
}
\label{rho-logT}
\end{figure}
%
In Fig.~\ref{rho-logT}, the temperature dependences of the in-plane resistivity $\rho_{a}$ are shown for single crystals of CeOBiS$_{2}$, LaOBiS$_{2}$, and LaO$_{0.5}$F$_{0.5}$BiS$_{2}$.
The $\rho_a (T)$ data of CeOBiS$_{2}$ and LaOBiS$_{2}$ show semiconducting behaviors and are three orders of magnitude larger than those of electron-doped LaO$_{0.5}$F$_{0.5}$BiS$_{2}$.
Such extremely weak conduction of CeOBiS$_{2}$ and LaOBiS$_{2}$ attests to the low impurity (or defect) concentration and high quality of the single crystals.
If the semiconducting behavior of CeOBiS$_{2}$ is analyzed tentatively using the activation-type relation $\rho_{a} = \rho_{0} \exp (\Delta/2k_{\rm B} T)$ [see Fig.~\ref{rho-logT} (b)], the energy gaps $\Delta_{\rm low}=0.93$ K below 5 K and $\Delta_{\rm high}=95$ K above 100 K are obtained; see Ref. 18 for a similar analysis of the semiconducting behavior observed in LaO$_{0.5}$F$_{0.5}$BiS$_{2}$ polycrystals.
Since these values are much smaller than the band gap energy of 0.4 eV obtained from a band structure calculation on LaOBiS$_{2}$~\cite{Usui_PRB_12}, the observed $\Delta_{\rm low}$ and $\Delta_{\rm high}$ may be attributed to minor impurity (or defect) conduction.
\\
\quad For the electron-doped LaO$_{0.5}$F$_{0.5}$BiS$_{2}$, $\rho_a (T)$ shows a superconducting transition at $T_{\rm c} \simeq 4$ K.
When $T > T_{\rm c}$, $\rho_a (T)$ shows weak but clear metallic behavior, in contrast to the semiconducting behavior observed in polycrystals~\cite{Awana_SSC_12}.
\\
\begin{figure}
\begin{center}
\includegraphics[width=\linewidth]{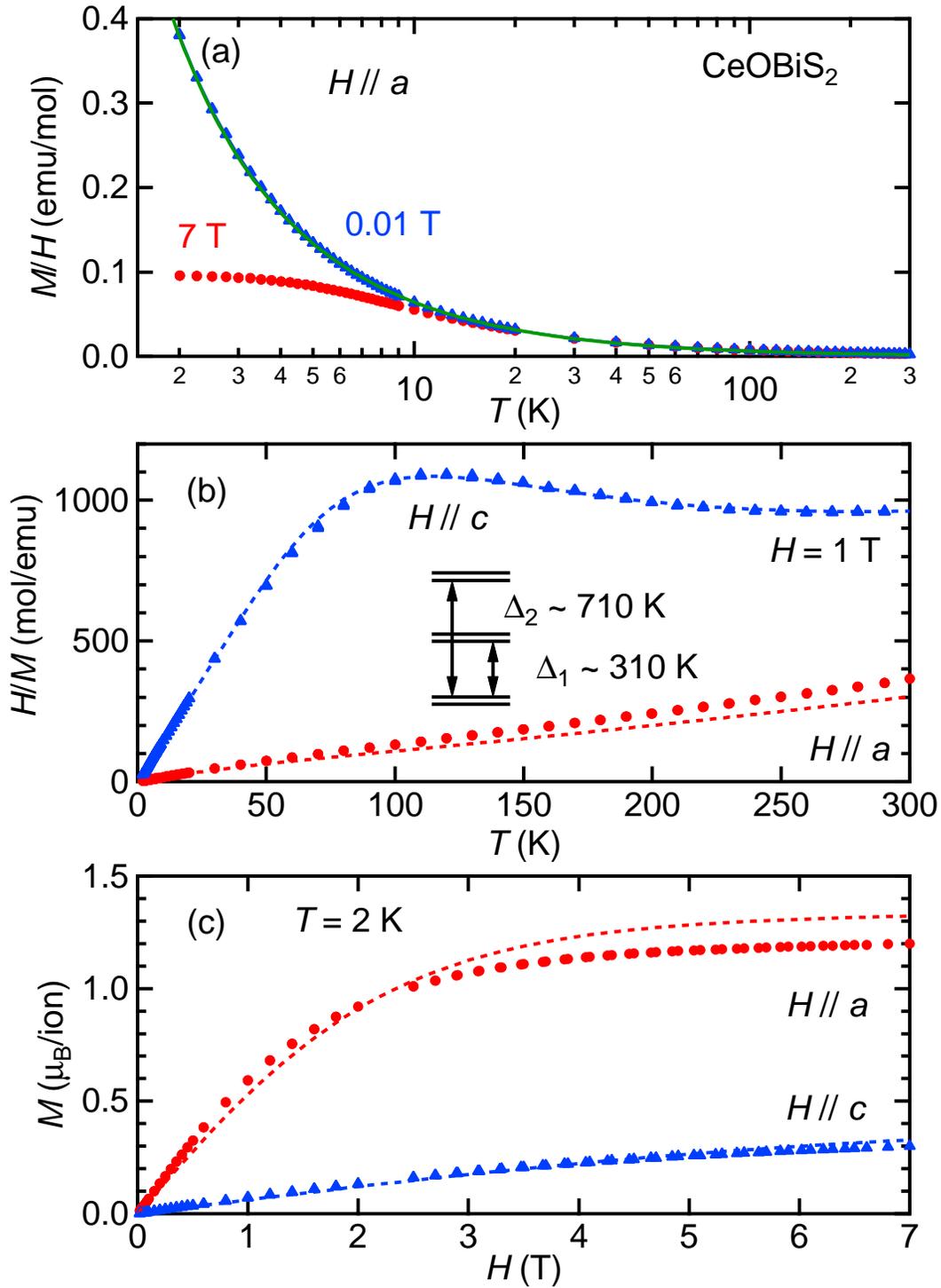}
\end{center}
\caption{(Color online) (a) Temperature dependences of DC susceptibility for single crystal of CeOBiS$_{2}$ measured at 0.01 and 7 T in $H$ // $a$.
The solid line represents the Curie Weiss fitting line below 20 K.
(b) Reciprocal susceptibility in $H$ // $a$ (solid circles) and $H$ // $c$ (solid triangles) measured at 1 T and (c) low-temperature magnetization in $H$ // $a$ and $H$ // $c$ measured at 2 K.
The determined CEF level scheme is shown in (b).
The broken lines are the calculated results of CEF analysis as described in the text.
}
\label{M-T}
\end{figure}
%
\quad Figure~\ref{M-T}(a) shows the temperature dependences of the DC susceptibility of CeOBiS$_{2}$ measured in 0.01 and 7 T for $H$ // $a$.
No anomaly indicating a phase transition is detected down to 2 K.
A Curie-Weiss fitting using $M/H = C/(T - \Theta_{\rm CW})$ to the data of 0.01 T below 20 K gives $C$ = 0.620(1) emu/K mol and $\Theta_{\rm CW}$ = 0.36(1) K.
The small value of $\Theta_{\rm CW}$ indicates the existence of weak ferromagnetic interactions among Ce ions when $M$ // $a$.
\\
\quad Figure ~\ref{M-T}(b) shows the temperature dependence of the reciprocal susceptibility $H/M$ measured in 1 T for $H$ // $a$ and $H$ // $c$.
The data demonstrate that Ce 4$f$-electron magnetic moments have significant magnetic anisotropy.
A shoulderlike anomaly appearing at approximately 100 K for $H$ // $c$ is a characteristic feature caused by CEF effects.
This fact indicates clearly that 4$f$-electrons are in a localized state and the valence of the Ce ions is 3+~\cite{valence}.
Figure~\ref{M-T}(c) shows the magnetic field dependence of magnetization at 2 K.
The larger values of $M$ for $H$ // $a$ compared with those for $H$ // $c$ indicate that the $ab$-plane is the magnetic easy plane.
\\
\quad Owing to CEF effects with the tetragonal point symmetry $C_{4v}$ ($4mm$), the $J$ = 5/2 multiplet of a Ce$^{3+}$ ion splits into three doublets.
The CEF Hamiltonian can be expressed as ${\mathcal H}_{\rm CEF} = B_{2}^{0} O_{2}^{0} + B_{4}^{0} O_{4}^{0} + B_{4}^{4} O_{4}^{4}$, where $O_{i}^{j}$ are Steven's operators and $B_{i}^{j}$ are CEF parameters~\cite{Stevens}.
A fitting of the CEF model to the $H/M (T)$ data was performed.
The best-fit CEF parameters are $B_{2}^{0}$ = 38.1(1) K, $B_{4}^{0}$ = -0.282(4) K, and $|B_{4}^{4}| \le$ 2 K, using which the model calculations of $\chi$ and $M$ are drawn in Figs.~\ref{M-T}(b) and \ref{M-T}(c).
Precise determination of $B_{4}^{4}$ (or the mixing parameter $\eta$ defined in Table II) is difficult since the behavior of $H/M (T)$ for $H$ // $c$ is insensitive to the $B_{4}^{4}$ term when $B_{4}^{4}$ is small as in the present case.
It is remarkable that the shoulderlike anomaly appearing in $H/M (T)$ for $H$ // $c$ is well reproduced by the CEF model, confirming our interpretation of localized 4$f$-electrons of Ce$^{3+}$ ions.
The CEF ground state is a pure $J_{z}$ = $\pm$1/2 doublet, and two excited doublets are located with the energy separations $\Delta_{1}$ = $\sim$ 310 K and $\Delta_{2}$ = $\sim$ 710 K.
The wave functions of each level are listed in Table II.
\\
\begin{table}[tb]
\caption{
CEF energy levels and the corresponding wave functions for CeOBiS$_{2}$.
The mixing parameter $\eta$ varies as $|\eta|$ $\textless$ 0.13 depending on $B_{4}^{4}$.
}
\label{t1}
\begin{center}
\begin{tabular}{ccc}
\hline
$E_{n}$ (K) & symmetry & wave functions \\
\hline
$\sim$ 710 & $\Gamma^{(2)}_{7}$ & $\sqrt{1-\eta^{2}}  \left| \pm \frac{5}{2} \right. \rangle + \eta \left| \mp \frac{3}{2} \right. \rangle$ \\
$\sim$ 310 & $\Gamma^{(1)}_{7}$ & $\eta \left| \pm \frac{5}{2} \right. \rangle - \sqrt{1-\eta^{2}} \left| \mp \frac{3}{2} \right. \rangle$ \\
0 & $\Gamma_{6}$ & $\left| \pm \frac{1}{2} \right. \rangle$ \\
\hline
\end{tabular}
\end{center}
\end{table}
\begin{figure}
\begin{center}
\includegraphics[width=\linewidth]{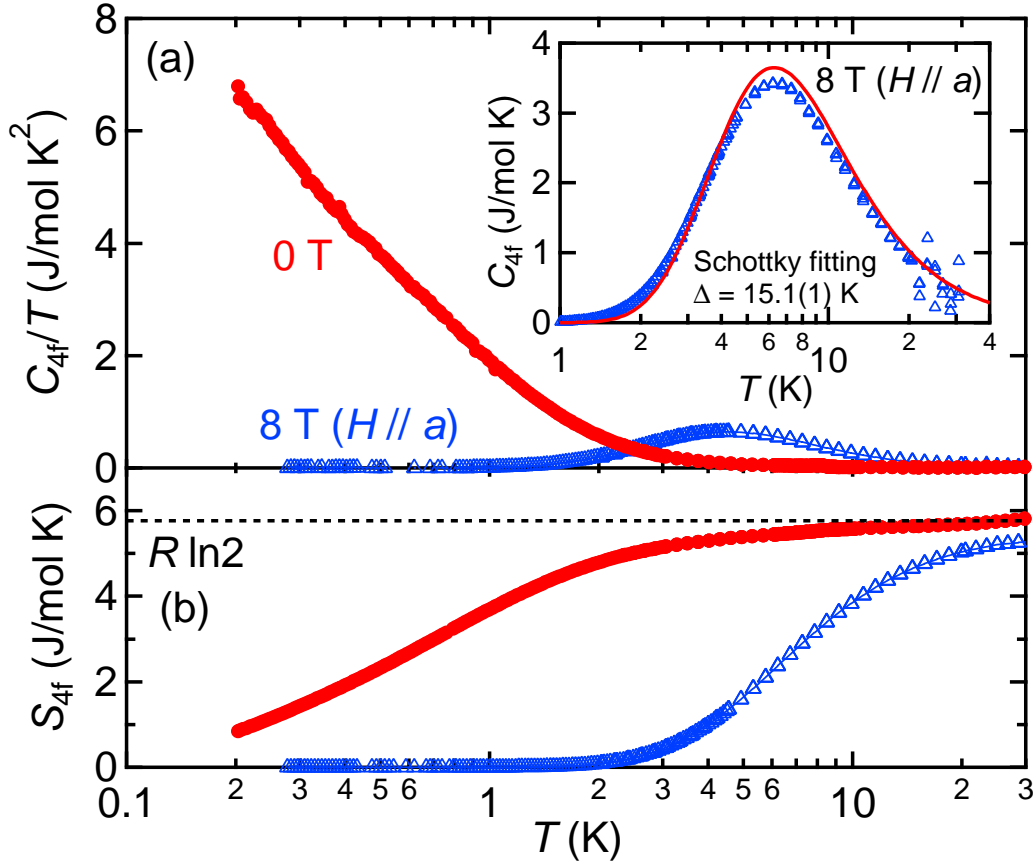}
\end{center}
\caption{(Color online) Temperature dependences of (a) $C_{4f}/T$ and (b) $S_{4f}$ for single crystals of CeOBiS$_{2}$ at 0 and 8 T in $H$ // $a$.
(a) Inset: Temperature dependence of $C_{4f}$ at 8 T in $H$ // $a$.
The solid line is the fitting curve using the two-level Schottky model. 
The absolute value of $S_{4f}$($T$, 0 T) is determined using the Maxwell relation $\partial M/\partial T |_{H} = \partial S/ \partial H |_{T}$ in combination with $S_{4f}$ (0.3 K, 8 T) = 0 and measured $M (T, H)$ data.
}
\label{CS4f-T}
\end{figure}
%
\quad The 4$f$-electron contribution to the specific heat $C_{4f}$ of CeOBiS$_{2}$ was obtained using $C_{4f} \equiv C_{{\rm CeOBiS}_{2}}-C_{{\rm LaOBiS}_{2}}$, where $C_{{\rm LaOBiS}_{2}}$ is the specific heat of LaOBiS$_{2}$ single crystals measured in 0 T.
The resulting data are plotted as $C_{4f}/T$ vs $\log T$ in Fig.~\ref{CS4f-T}(a).
No anomaly indicating a phase transition was detected down to 0.2 K in any field, indicating that 4$f$-electrons are in a paramagnetic state in the measured $T$ and $H$ regions.
The most salient feature in Fig.~\ref{CS4f-T}(a) is that $C_{4f}/T$ in zero field continues to increase with decreasing temperature below $\sim$ 3 K~\cite{Kondo}.
It shows distinct $- \log T$ dependence for about one decade in temperature.
\\
\quad The 4$f$-electron contribution to the entropy $S_{4f}$ calculated using the $C_{4f}$ data is shown in Fig.~\ref{CS4f-T}(b) as a function of $\log T$.
In the temperature range of around $5-30$ K, $S_{4f}$ shows plateau behavior with a value of $R \ln 2$, consistent with the CEF level scheme obtained above.
From this fact, it is obvious that the $- \log T$ dependence of $C_{4f}/T$ originates in the $J_{z}$ = $\pm$1/2 ground-state doublet.
The characteristic temperature $T^*$ of the $- \log T$ behavior can be estimated from the fitting using $C_{4f}/T \propto \log (T/T^{*})$, which gives $T^*$= 1.8(1) K.
Therefore, a measurement of the magnetic susceptibility below $T^{*}$ is needed to investigate the magnetic properties related to the $- \log T$ behavior; in Fig.~\ref{M-T}(a), the deviation from the Curie-Weiss behavior is hardly noticeable above 2 K.
\\
\quad By applying fields, the $- \log T$ divergence of $C_{4f}/T$ is suppressed and it shifts to higher temperatures with its structure changing.
Eventually in 8 T, a Schottky peak structure is observed at approximately 6 K.
This peak is attributable to the thermal excitation in the Zeeman-split ground-state doublet, reflecting the fact that quantum fluctuations in 4$f$ magnetic moments dominating in zero field are significantly suppressed in 8 T.
The peak height of 3.60 J/K mol, which is slightly smaller than 3.65 J/K mol expected for a two-level system, reflects the broadening of the peak probably caused by remaining quantum fluctuations.
The Zeeman energy separation estimated from the peak position is 15.1 K, which is slightly larger than 13.8 K calculated for the CEF ground state; this indicates the existence of weak ferromagnetic interactions among Ce ions.
\\
\quad Despite the nearest-neighbor (nn) Ce-Ce distance being only 3.75 $\AA$, no magnetic ordering occurs down to 0.2 K.
This is attributable to several characteristic features of the crystal structure (see Fig.~\ref{Crystal_Structure}).
CeO block layers are separated from each other by a large distance of $c$ (= 13.49 $\AA$), and two BiS$_2$ layers weakly bound by van der Waals forces are inserted in between\cite{Liu_condmat_14}.
This feature probably makes CeO interlayer interactions extremely weak, making this material a nearly 2D system (note that, in electron-doped isomorphic Nd(OF)BiS$_2$, the 2D nature of the Fermi surface topology has also been confirmed by Angle-Resolved Photoemission Spectroscopy measurements~\cite{Ye_PRB_14}).
\\
\quad Magnetic interactions in a CeO block layer are probably mediated by superexchange interactions.
In a CeO layer, Ce ions form an undulating ichimatsu-patterned (checkerboard-patterned) square lattice, which consists of two square sublattices A and B intervening with each other.
There are two types of Ce-O-Ce bond and each provides different interactions, $J_1$ (inter-sublattice) and $J_2$ (intra-sublattice).
Because these bonds have the same total bond length of 4.69 $\AA$ and slightly different bond angles of 106 and 116$^{\circ}$, $J_1$ and $J_2$ may compete with each other, causing a CeO layer to become a frustrated magnet. \cite{Comment}
In addition, the 4$f$-electron spatial charge distribution of the $J_{z}$ = $\pm$1/2 ground state is elongated along the $c$-axis\cite{Walter_1986}.
Therefore, it is expected that $J_{1}$ and $J_{2}$ each have sufficiently low strength. 
\\
\quad The internal degrees of freedom that the $J_{z}$ = $\pm$1/2 ground state have are magnetic moments of ${J_x, J_y}$ $(\Gamma_5)$, and ${J_z}$ $(\Gamma_2)$.
Therefore, interactions among Ce ions can be expressed with the spin-1/2 XXZ model.
In a 2D quantum spin-1/2 XXZ system along with the $J_1-J_2$ frustration, theoretical calculations show that no long-range order (LRO) appears in a certain parameter range called the ``quantum paramagnetic phase'' \cite{Bishop_2008, Misguich_2013}.
Therefore, it may be possible that CeOBiS$_2$ is located in this parameter region (or in the vicinity of the QCPs existing around it), where quantum fluctuations are expected to dominate at low temperatures close to $T=0$.
\\
\quad Considering that $\Theta_{\rm CW}$ = + 0.36 K reflects the ferromagnetic interaction among Ce ions at the wave vector ${\mib q}$ = 0, the fact that $S_{4f}$ starts to decrease below approximately 4 K indicates the existence of dominant antiferromagnetic (${\mib q} \neq 0$) components with the characteristic energy of about 4 K, which causes the $- \log T$ divergence of $C_{4f}/T$.
The lack of magnetic ordering down to 0.2 K, which is more than one order of magnitude smaller than 4 K, is in agreement with the frustration scenario.
\\
\quad Among the geometrically frustrated nonmetallic magnets with no LRO, many have been discussed as candidate of ``quantum spin liquids" \cite{Balents_Nature_2010} (e.g., NiGa$_{2}$S$_{4}$\cite{Nakatsuji_Science_2005} and Na$_{4}$Ir$_{3}$O$_{8}$ \cite{Okamoto_PRL_2007}). 
These materials show roughly $T^{2}$-dependent specific heat at low temperatures, and no compounds that show exotic $-\log T$ dependence except for CeOBiS$_{2}$ have been reported in the literature. 
Therefore, the present findings strongly indicate that CeOBiS$_{2}$ is the first material found to be located at a QCP.
\begin{acknowledgments}
We are grateful to W. Fujita for the single-crystal X-ray analysis and to T. Hotta, M. Nohara, H. Otsuka, and R. Shiina for fruitful discussions. 
This work was supported by a Grant-in-Aid for Young Scientists (B) (No. 24740239) from JSPS, SICORP-EU-Japan, and JST-ALCA.
\end{acknowledgments}


\begin{thebibliography}{9}
\bibitem{Lohneysen_2007} H. v. Lohneysen, A. Rosch, M. Vojta, and P. W\"{o}lfle, Rev. Mod. Phys. {\bf 79}, 1015 (2007).
\bibitem{Stewart_2001} G. R. Stewart, Rev. Mod. Phys. {\bf 73}, 797 (2001); {\bf 78}, 743 (2006).
\bibitem{Steglich_1996} F. Steglich, B. Buschinger, P. Gegenwart, M. Lohmann, R. Helfrich, C. Langhammer, P. Hellmann, L. Donnevert, S. Thomas, A. Link, C. Geibel, M. Lang, G. Sparn, and W. Assmus, J. Phys.: Condens. Matter {\bf 8}, 9909 (1996).
\bibitem{Matsumoto_2011} Y. Matsumoto, S. Nakatsuji, K. Kuga, Y. Karaki, N. Norie, Y. Shimura, T. Sakakibara, A. H. Nevidomskyy, and P. Coleman, Science {\bf 331}, 316 (2011).
\bibitem{Mizuguchi_PRB_12} Y. Mizuguchi, H. Fujihisa, Y. Gotoh, K.  Suzuki, H. Usui, K. Kuroki, S. Demura, Y. Takano, H. Izawa, and O. Miura, Phys. Rev. B {\bf86}, 220510(R) (2012).
\bibitem{Mizuguchi_JPSJ_12} Y. Mizuguchi, S. Demura, K. Deguchi, Y. Takano, H. Fujihisa, Y. Gotoh, H. Izawa, and O. Miura, J. Phys. Soc. Jpn. {\bf81}, 114725 (2012).
\bibitem{Singh_JACS_12} S. K. Singh, A. Kumar, B. Gahtori, S. Kirtan, G. Sharma, S. Patnaik, and V. P. S. Awana, J. Am. Chem. Soc. {\bf134}, 16504 (2012).
\bibitem{Demura_condmat_12} S. Demura, Y. Mizuguchi, K. Deguchi, H. Okazaki, H. Hara, T. Watanabe, S.J. Denholme, M. Fujioka, T. Ozaki, H. Fujihisa, Y. Gotoh, O. Miura, T. Yamaguchi, H. Takeya, and Y. Takano, J. Phys. Soc. Jpn. {\bf82}, 033708 (2013). 
\bibitem{VESTA} K. Momma and F. Izumi, J. Appl. Crystallogr. {\bf 44}, 1272 (2011).
\bibitem{Wolowiec_PRB_13} C. T. Wolowiec, D. Yazici, B. D. White, K. Huang, and M. B. Maple, Phys. Rev. B {\bf 88}, 064503 (2013).
\bibitem{Cuprate_Review} P. A. Lee, N. Nagaosa, and X. G. Wen, Rev. Mod. Phys. {\bf78}, 17 (2006).
\bibitem{Kamihara_JACS_08} Y. Kamihara, T. Watanabe, M. Hirano, and H. Hosono, J. Am. Chem. Soc. {\bf130}, 3296 (2008).
\bibitem{Pnictide_Review} K. Ishida, Y. Nakai, and H. Hosono, J. Phys. Soc. Jpn. {\bf 78}, 062001 (2009).
\bibitem{Nagao_SSC_14} M. Nagao, A. Miura, S. Demura, K. Deguchi, S. Watanuki, T. Takei, Y. Takano, N. Kumada, and I. Tanaka, Solid State. Commun. {\bf178}, 33 (2014).
\bibitem{Xing_PRB_12}J. Xing, S. Li, X. Ding, H. Yang, and H.-H. Wen, Phys. Rev. B {\bf 86}, 214518 (2012).
\bibitem{Demura_condmat_13}S. Demura, K. Deguchi, Y. Mizuguchi, K. Sato, R. Honjyo, A. Yamashita, T. Yamaki, H. Hara, T. Watanabe, S. J. Denholme, M. Fujioka, H. Okazaki, T. Ozaki, O. Miura, T. Yamaguchi, H. Takeya, and Y. Takano, arXiv:1311.4267.
\bibitem{SHLEX-97} G. M. Sheldrick, Acta Crystallogr. A {\bf 64} 112 (2008).
\bibitem{Kotegawa_JPSJ_12} H. Kotegawa, Y. Tomita, H. Tou, H. Izawa, Y. Mizuguchi, O. Miura, S. Demura, K. Deguchi, and Y. Takano, J. Phys. Soc. Jpn. {\bf81}, 103702 (2012).
\bibitem{Usui_PRB_12} H. Usui, K. Suzuki, and K. Kuroki, Phys. Rev. B {\bf 86}, 220501(R) (2012).
\bibitem{Awana_SSC_12} V. P. S Awana, A. Kumar, R. Jha, S. Kumar, J. Kumar, A. Pal, Shruti, J. Saha, and S. Patnaik, Solid State Commun. {\bf 157}, 21 (2012).
\bibitem{valence} Recent X-ray absorption spectroscopy (XAS) measurement on polycrystalline CeOBiS$_{2}$ has shown that the Ce-ion valence is 3.15 at room temperature, indicating that Ce ions are in a mixed valence state~\cite{Sugimoto_PRB_14}. This result does not seem to be consistent with our present data on single crystals. To resolve this discrepancy, XAS measurements on our single crystals are under way.
\bibitem{Sugimoto_PRB_14} T. Sugimoto, B. Joseph, E. Paris, A. Iadecola, T. Mizokawa, S. Demura, Y. Mizuguchi, Y. Takano, and N. L. Saini, Phys. Rev. B {\bf 89}, 201117(R) (2014). 
\bibitem{Stevens} K. W. H. Stevens: Proc. Phys. Soc., Sect. A {\bf 65}, 209 (1952).
\bibitem{Kondo} If the $- \log  T$ behavior of $C_{4f}(T)$ is compared tentatively with the exact solution of the spin 1/2 impurity Kondo model~\cite{Tsvelick_1983}, the Kondo temperature $T_{\rm K}$ of $\sim 1$ K is estimated, even though some deviation remains between the data and the model curve.
\bibitem{Tsvelick_1983} A. M. Tsvelick and P. B. Wiegmann, Adv. Phys. {\bf 32}, 453 (1983).
\bibitem{Liu_condmat_14} Q. Liu, X. Zhang, H. Jin, K. Lam, J. Im, A. J. Freeman, and A. Zunger, arXiv:1408.6004.
\bibitem{Ye_PRB_14} Z. R. Ye, H. F. Yang, D. W. Shen, J. Jiang, X. H. Niu, D.L. Feng, Y. P. Du, X. G. Wan, J. Z. Liu, X. Y. Zhu, H. H. Wen, and M. H. Jaing, Phys. Rev. B {\bf 90}, 045116 (2014).
\bibitem{Comment} Even when the dominant interactions are mediated via Ce-S(2)-Ce bond, the same discussion holds. In this case, the total bond length is 6.20 \AA~and the bond angles are 80.4 and 130.8$^{\circ}$ for $J_{1}$ and $J_{2}$, respectively.
\bibitem{Walter_1986} U. Walter, Z. Phys. B {\bf 62}, 299 (1986).
\bibitem{Misguich_2013} G. Misguich and C. Lhuillier: in $Frustrated$ $spin$ $systems$, ed. H. T. Diep (World-Scientific, Singapore, 2013) 2nd ed., Chap. 5.
\bibitem{Bishop_2008} R. F. Bishop, P. H. Y. Li, R. Darradi, J. Schulenburg, and J. Richter, Phys. Rev. B {\bf 78}, 054412 (2008).
\bibitem{Balents_Nature_2010} L. Balents, Nature {\bf 464}, 199 (2010).
\bibitem{Nakatsuji_Science_2005} S. Nakatsuji, Y. Nambu, H. Tonomura, O. Sakai, S. Jonas, C. Broholm, H. Tsunetsugu, Y. Qiu, and Y. Maeno, Science {\bf 309}, 1697 (2005).
\bibitem{Okamoto_PRL_2007} Y. Okamoto, M. Nohara, H. A. Katori, and H. Takagi, Phys. Rev. Lett. {\bf 99}, 137207 (2007).
\end{thebibliography}
\end{document}